\documentclass[12pt]{article}
\usepackage{epsfig}
\usepackage{amsmath}
\usepackage{amsfonts}

\begin{document}
\pagestyle{empty}
\begin{flushright}
UMN-TH-2425/05\\
December 2005
\end{flushright}
\vspace*{5mm}

\begin{center}
{\Large\bf Localized U(1) Gauge Fields, Millicharged\\
Particles, and Holography}
\vspace{1.0cm}

{\sc Brian Batell}\footnote{E-mail:  batell@physics.umn.edu}
{\small and} 
{\sc Tony Gherghetta}\footnote{E-mail:  tgher@physics.umn.edu}
\\
\vspace{.5cm}
{\it\small {School of Physics and Astronomy\\
University of Minnesota\\
Minneapolis, MN 55455, USA}}\\
\end{center}

\vspace{1cm}
\begin{abstract}
We consider U(1) gauge fields in a slice of AdS$_5$ with bulk and 
boundary mass parameters. The zero mode of a bulk U(1) gauge field can be 
localized either on the UV or IR brane. This leads to a simple model 
of millicharged particles in which fermions can have arbitrarily small 
electric charge. In the electroweak sector we also discuss phenomenological 
implications of a localized U(1)$_Y$ gauge boson. 
Using the AdS/CFT correspondence we present the 4D holographic interpretation 
of the 5D model. In particular the photon is shown to be a composite particle 
when localized near the IR brane, whereas it is elementary when localized near 
the UV brane. In the dual interpretation the ``millicharge'' results from 
an elementary fermion coupling to a composite photon via a vector current with 
large anomalous dimension.

\end{abstract}

\vfill
\begin{flushleft}
\end{flushleft}
\eject
\pagestyle{empty}
\setcounter{page}{1}
\setcounter{footnote}{0}
\pagestyle{plain}

\section{Introduction}

The AdS/CFT correspondence~\cite{malda} has provided a simple and compelling 
framework in which to study four dimensional (4D) gauge theories. Gauge 
fields propagating 
in the background of an AdS$_5$ warped geometry provide a weak-coupling 
description of the nontrivial dynamics that occurs in strongly-coupled 
4D gauge theories. In a slice of AdS$_5$~\cite{rs} the best known example 
is a massless U(1) gauge field~\cite{gaugeb1,gaugeb2}. The zero mode of 
this bulk field is not localized and has a flat profile. In the dual
4D interpretation this corresponds to a part-elementary and part-composite 
photon eigenstate which resembles $\gamma-\rho$ mixing in QCD~\cite{arp}.

It is perhaps not so well known that the zero mode U(1) gauge field can 
in fact be localized anywhere in the warped bulk by adding both bulk and 
brane localized mass terms~\cite{gbl1,gbl2}. Essentially, as first pointed 
out for bulk scalar fields~\cite{gp1}, the brane localized mass terms serve 
to alter the boundary conditions in such a way that a zero mode solution 
is still allowed. Although this requires a fine tuning between bulk and 
boundary mass parameters the zero mode photon can be 
localized anywhere in the warped bulk.

Several interesting phenomenological scenarios are then possible. Just as 
separating bulk fermions from the Higgs boson in the warped bulk can lead to 
Yukawa coupling hierarchies~\cite{gn,gp1}, a simple model of millicharged 
particles by separating fermions in the warped bulk from a localized U(1) 
gauge boson can also be constructed. This will allow fermions to have 
arbitrarily small electric charge and is a new way to obtain 
millicharged particles. Moreover a grand unified scenario can be considered
in the warped bulk by generating kinetic mixing between two U(1) gauge 
fields~\cite{holdom,as}. 
In addition a localized U(1) gauge boson will lead to different possibilities 
in the electroweak sector from those considered so far in warped 
Randall-Sundrum models. In particular stringent 
electroweak constraints from bulk Abelian gauge fields can be avoided 
without delocalizing fermions or adding brane kinetic terms.

The most interesting aspect of the localized bulk U(1) gauge field is
that it can be given a 4D holographic description. Much like previous 
analyses for bulk fermions~\cite{cp} and 
bulk gravitons~\cite{gpp}, the UV (IR) brane localized gauge fields can be 
shown to correspond to elementary (composite) photon eigenstates in the dual 
4D theory. The boundary mass provides a continuous parameter which connects 
these two limiting possibilities. In particular when the
photon zero mode is localized on the IR brane this corresponds to a composite
photon in the dual 4D theory. This is an example of emergent behavior since
the photon only exists at large distance scales. The dual holographic 
description then allows us to interpret millicharged particles as resulting 
from elementary fermions coupling to a composite photon via a 
vector current with large anomalous dimension. 

The outline of this paper is as follows. In Section 2 we will review
Abelian gauge fields in warped space. The equations of motion for both 
massless and massive gauge fields have already been studied. However
dealing with gauge fields in warped extra dimensions can be tricky because of 
additional scalar fields that arise in the dimensional reduction, and the 
fate of these modes is often obscured. Instead we will employ a five 
dimensional (5D) St\"{u}ckelberg mechanism which maintains 
manifest gauge invariance. This provides a simple way to identify the 
dynamical scalar fields of the theory, while still being able to decouple the 
photon and preserve 4D gauge invariance. The phenomenological applications 
of localized U(1) gauge fields are then presented in Section 3. This
includes millicharged particles, and a model of the electroweak sector.
In Section 4 the holographic interpretation of the bulk 5D model is 
presented, including the dual interpretation of millicharged particles.
Concluding remarks are given in Section 5.

\section{Abelian gauge fields in warped space}

\subsection{Massless bulk gauge fields}

In a slice of AdS$_5$, the metric is~\cite{rs}
\begin{equation}
ds^2=e^{-2 ky}\eta_{\mu\nu}dx^{\mu} dx^{\nu}+dy^2,
\end{equation}
where $k$ is the AdS curvature scale. The extra coordinate $y$ ranges from 
$0<y< \pi R$. At the boundaries $y=0$ and $y=\pi R$ there exist two 
three-branes, called the ultraviolet (UV) and infrared (IR) brane, 
respectively. We label 5D coordinates with Latin indices ($A,B,$...) and 
4D coordinates with Greek indices ($\mu,\nu$,...). Greek indices are raised 
and lowered strictly with the 4D Minkowski metric, $\eta_{\mu\nu}$, which has 
signature $(-,+,+,+)$.

Before we examine massive vector fields and the possibility of localizing the 
zero mode, it will be instructive to review the massless case and compare our 
approach to results obtained previously~\cite{gaugeb1,gaugeb2,rsch}. The 5D action for a massless 
gauge field in warped space is given by
\begin{equation} 
S=\int d^5x\sqrt{-g}\left(-\frac{1}{4}g^{MN}g^{RS}F_{MR}F_{NS}\right).
\label{masslessaction}
\end{equation}
Rather than choosing a particular gauge to simplify the analysis, our 
strategy will be to write the action in terms of gauge invariant combinations 
of the fields. To this end, we parameterize the 5D vector $A_M$ in the 
following way:
\begin{equation}
A_M=( \widehat{A}_\mu+\partial_\mu \varphi, A_5)
\label{par}
\end{equation}
where $\partial_\mu \widehat{A}^\mu=0$. We interpret $\widehat{A}_\mu$ and 
$\varphi$ as the transverse and longitudinal components of $A_\mu$, 
respectively. Under the gauge transformation 
$A_M\rightarrow A_M+ \partial_M\lambda$, with gauge parameter $\lambda$, 
the transverse vector $\widehat{A}_\mu$ is invariant, while the scalar 
longitudinal mode transforms 
as $\varphi \rightarrow \varphi+\lambda$. Three degrees of freedom  
are contained in $\widehat{A}_\mu$ while the scalars $\varphi$ and $A_5$ 
account for another two, making up a total of five degrees of freedom for 
the 5D vector field.
However, the dynamics and the gauge symmetry imply that there are only three 
physical degrees of freedom, and this fact will guide our analysis. 

With this parameterization, the action (\ref{masslessaction}) becomes
\begin{equation}
S=\int d^5x \left[ -\frac{1}{4}\widehat{F}^2_{\mu\nu}-\frac{1}{2}e^{-2ky}
(\partial_5 \widehat{A}_\mu)^2 -\frac{1}{2}e^{-2ky}(\partial_\mu \psi)^2 
\right]~,
\label{masslessaction2}
\end{equation}
where we have defined the gauge invariant field $\psi=A_5-\varphi'$, with
prime ($'$) denoting differentiation with respect to $y$. Note that the 
action (\ref{masslessaction2}) has decoupled in terms of the fields 
${\widehat A}_\mu$ and $\psi$. Performing a Kaluza-Klein decomposition of the 
vector, we find the standard result of a single massless mode (two degrees of 
freedom) and a tower of Proca fields (three degrees of 
freedom)~\cite{gaugeb1,gaugeb2}. We therefore expect that the bulk dynamics 
will allow only a single massless scalar mode. Indeed, the equation of motion 
for $\psi$ is simply $\Box \, \psi(x,y)=0$, which means that there exists 
only a massless mode, 
\begin{equation}
\psi(x,y)=\psi^0(x) f^0(y).
\end{equation}
Hence, we have found our remaining degree of freedom at the massless level. 

Although we have found a massless scalar mode, the wavefunction of $\psi$ is 
undetermined from the action. Instead to find the wavefunction, we can start 
from the 5D equation of motion, which comes from varying the action 
(\ref{masslessaction}).
This leads to the first order differential equation
\begin{equation}
\partial_5\left(e^{-2ky} \psi(x,y)\right)=0~.
\label{psi1}
\end{equation}
The solution is given by
\begin{equation}
\psi(x,y)=C \psi^0(x) e^{2ky},
\end{equation}
where $C$ is a normalization constant. Substituting the solution back into 
the action, gives
\begin{equation}
S=-C^2 \int d^4x\, \frac{1}{2}(\partial_\mu \psi)^2 \int dy \, e^{2ky}
+ \dots~,
\end{equation}
and therefore the profile of $\psi$ with respect to a flat metric is given by 
\begin{equation}
f^0(y) \propto e^{ky}.
\label{psiprofile}
\end{equation}

It is common when dealing with 5D gauge fields to expand the scalar field $A_5$ in terms of a derivative of a function \cite{rsch}, $A^n_5(x)\partial_5 f^n(y)$. This has the virtue of diagonalizing the interaction terms between $A^n_\mu$ and $A^n_5$. Since $A_5$ is contained in the gauge invariant field $\psi$, it is natural to ask what happens if we expand the field $\psi(x,y)$ in terms of a derivative of a function. From (\ref{psi1}), it is clear that the equation of motion is second order and we get two solutions. One solution is a constant $c$, meaning that the zero mode vanishes: $\partial_5 c=0$. The other solution gives us the same bulk profile as (\ref{psiprofile}), so clearly the physics is the same regardless of which way the scalar field is expanded. Whether or not the mode actually exists depends on the boundary conditions of the theory. 

Another approach is to add the following gauge fixing term to the 
Lagrangian~\cite{rsch,pgb}:
\begin{equation}
      {\cal L}_{GF}=-\frac{1}{2 \xi}(\partial_\mu A^\mu+\xi 
         \partial_5(e^{-2ky}A_5))^2~,
\end{equation}
which removes the interaction term between $A_\mu$ and $A_5$. With 
this choice of gauge, the equation of motion for $A_5$ becomes
\begin{equation}
\left( e^{2 ky} \Box +\xi e^{2ky} \partial_5^2 e^{-2ky} \right)A_5(x,y)=0~.
\end{equation}
Performing a Kaluza-Klein decomposition, we find that the wavefunction of the massive modes depends on the gauge parameter $\xi$, indicating that these are fictitious degrees of freedom. This agrees with our analysis in which we found no massive scalar modes.  However, since $\Box A_5^0(x)=0$, the equation of motion for the zero mode is independent of $\xi$, which means that the massless mode is indeed a physical particle. Its profile with respect to a flat background agrees with (\ref{psiprofile}).

Working with gauge invariant fields allows us to clearly identify the 
dynamical degrees of freedom contained in $A_M$. We have seen that our 
analysis of the massless gauge field is consistent with other approaches. 
This approach will be even more helpful when we examine massive bulk vector 
fields.
  
\subsection{Massive bulk gauge fields}

We now turn to the study of massive gauge fields. Essentially, our analysis 
will follow~\cite{gbl1}, but we will demand that 5D gauge invariance be a 
symmetry of our theory. As we will see, this allows us to cleanly identify 
the scalar degrees of freedom in the theory. 
One way to restore gauge invariance to the theory of the massive gauge field 
is to add a degree of freedom by simply promoting the gauge parameter to a 
dynamical field. This is the famous St\"{u}ckelberg formalism 
(for a review, see Ref~\cite{stuckreview}). 
The action is
\begin{equation} S=\int
d^5x\sqrt{-g}\left(-\frac{1}{4}g^{MN}g^{RS}F_{MR}F_{NS}
     -\frac{1}{2}(\partial_M\phi - m A_M)^2 \right),
\label{stu}
\end{equation}
which is invariant under the gauge transformation, 
$A_M\rightarrow A_M+\partial_M\lambda$ provided the field $\phi$ 
simultaneously transforms as $\phi\rightarrow \phi+m\lambda$. Following 
Ref~\cite{gbl1}, we will parameterize $A_M$ as in (\ref{par}).
There are three degrees of freedom contained in the scalar fields 
$\varphi$, $\phi$, and $A_5$, while $\widehat{A}_\mu$ contains three degrees 
of freedom. This makes a total of six degrees of freedom for the 
St\"{u}ckelberg action (\ref{stu}). Of course, as in the massless case, 
two degrees of freedom are removed by gauge invariance and dynamics.

Rewriting the action using our parameterization for $A_M$ (\ref{par}), we 
again find that the transverse vector $\widehat{A}_\mu$ decouples from the 
scalar fields. The action for $\widehat{A}_\mu$ becomes
\begin{equation}
S({\widehat A}_\mu)=\int d^5 x\left[ -\frac{1}{4}\widehat{F}^2_{\mu\nu}
-\frac{1}{2}e^{-2ky}(\partial_5 \widehat{A}_\mu)^2-\frac{1}{2}e^{-2ky}m^2 
\widehat{A}_\mu^2\right]~.
\label{vector}
\end{equation}
This will be our starting point in the next section. We will see that by 
adding a specific boundary mass term, it is possible to localize the zero 
mode of $\widehat{A}_\mu$. 

Before that let us analyze the scalar modes. The action for the scalar 
fields, which follows from (\ref{stu}), can be written in the form:
\begin{eqnarray}
S(\varphi,\phi,A_5) &=&
   \int d^5x  \left[ -\frac{1}{2}e^{-2ky}\left(\partial_\mu 
(A_5-\varphi')\right)^2
-\frac{1}{2}e^{-2ky}\left(\partial_\mu (\phi- m\varphi)\right)^2
\right. \nonumber \\ 
& &\qquad\qquad\qquad\qquad \left.-\frac{1}{2}m^2 e^{-4ky}(A_5-\frac{1}{m}
\phi')^2 \right]~.
\label{scalar}
\end{eqnarray}
Notice that this action is gauge invariant, which indicates that we really only have two degrees of freedom. Varying the action (\ref{scalar}), we get a system of coupled differential equations in terms of the fields $A_5, \varphi$, and $\phi$. To isolate the true dynamical variables, let us define the following gauge invariant scalar fields:
\begin{eqnarray}
\psi &= & A_5-\varphi'~, \nonumber \\
\rho & = & \phi- m \varphi~, \nonumber \\
\sigma & = & A_5-\frac{1}{m}\phi'~.
\label{ginv}
\end{eqnarray}
The action (\ref{scalar}) can then be written solely in terms of these fields.
However these fields are not independent, since $\psi-\rho'/m=\sigma$. If we
eliminate $\rho$ in favor of $\psi$ and $\sigma$ then the equations of motion 
are:
\begin{eqnarray}
& \Box(e^{2ky}\Box +\partial_5^2 -2k\partial_5-m^2) \psi(x,y) =0~,
\label{psieq}\\
& (e^{2ky}\Box +\partial_5^2 -6k\partial_5 + 8k^2-m^2) \sigma(x,y) =0~.
\label{sigmaeq}
\end{eqnarray}
The equation for $\sigma$ is identical to the equation for $A_5$ found in 
Ref.~\cite{gbl1}. This is consistent since working in the gauge $\phi=0$ 
corresponds to $\sigma=A_5$ (\ref{ginv}). It is clear from (\ref{ginv}), (\ref{psieq}), and (\ref{sigmaeq}) that the equations of motion for $A_5$, $\varphi$, and $\phi$ are indeed dependent on the choice of gauge. This underscores the advantage of working with gauge invariant fields.  

Note that the bulk equations do permit two massless scalar modes, 
$\sigma^0(x)$ and $\psi^0(x)$. From a phenomenological standpoint, these 
modes are usually undesirable because they are ruled out experimentally. 
However, boundary conditions can be imposed so that the zero modes vanish, 
and we will therefore not consider the phenomenological implications further 
in this paper. 

One additional point deserves to be mentioned. Our approach was to define 
gauge invariant combinations of the fields by separating the vector and 
scalar fields contained in $A_M$. In practice, this is equivalent to choosing 
a gauge $\partial_\mu A^\mu =0$. 
Of course, the analysis can be done in another gauge. However, from a 
physical standpoint, choosing to work in a particular gauge obscures the 
dynamics. The 5D equation of motion is dependent on the gauge choice. We know 
physically that the wavefunction of the true dynamical fields should not 
depend on the gauge, and working with gauge invariant fields allows us to 
avoid this problem. 

\subsection{Localizing the photon}
Let us now return to the transverse vector modes $\widehat{A}_\mu$.
Varying the action (\ref{vector}), we find the equations of motion for 
$\widehat{A}_\mu$:
\begin{equation}
(e^{2ky}\Box +\partial_5^2 -2k\partial_5-a k ^2) \widehat{A}_\mu(x,y) =0~,
\label{theeom}
\end{equation}
where we have defined the bulk mass $m^2=ak^2$ with dimensionless parameter
$a$. To perform the Kaluza-Klein decomposition, we expand $\widehat{A}_\mu$ 
into eigenmodes
\begin{equation}
\widehat{A}_\mu(x,y)=\sum_{n=0}^{\infty} \widehat{A}^{n}_\mu(x)f^{n}(y)~,
\label{expand}
\end{equation}
where $f^{n}$ satisfies
\begin{equation}
   (\partial^2_5 -2k\partial_5 -a k^2+e^{2ky}m^2_n)f^{n}(y)=0~,
\label{eomy}
\end{equation}
and obeys the orthonormal condition
\begin{equation}
\int_0^{\pi R} dy\, f^nf^m=\delta^{nm}~.
\label{norm}
\end{equation}
The solution for the zero mode ($m_0=0$) is
\begin{equation}
f^{0}(y)=C_1~e^{(1+\sqrt{1+a})ky}+ C_2~e^{(1-\sqrt{1+a})ky}~.
\end{equation}
For arbitrary boundary conditions one finds that $C_1=C_2=0$, and therefore
there is no zero mode. However, as for a bulk scalar field~\cite{gp1},
consider adding the following boundary mass term to the 
action~\cite{gbl1,gbl2}:
\begin{equation}
S_{bdy}= -\int d^5 x \sqrt{-g}~  \alpha\, k ~ g^{\mu\nu}
A_\mu A_\nu \left( \delta(y) -\delta(y-\pi R) \right)~,
\end{equation}
where $\alpha$ is a dimensionless parameter. Note that we have chosen equal 
and opposite boundary mass terms. This brane-localized mass term alters 
the boundary conditions for $\widehat{A}_\mu$, which become
\begin{equation}
\left(\partial _5 \widehat{A}_\mu -\alpha\,k \widehat{A}_\mu \right) 
   \bigg\vert_{0,\pi R} =0~.
\end{equation}
For generic values of $\alpha$, there is again no zero mode allowed. However, 
if the bulk and boundary mass parameters are tuned in the following way:
\begin{equation}
  \alpha_\pm=1\pm \sqrt{1+a}~,
\label{alpha}
\end{equation}
then either $C_1$ or $C_2$ is non-vanishing. Under this condition 
(\ref{alpha}), there is a normalizable massless mode solution. 

We can consider both the $\alpha_+$ and the $\alpha_-$ branches 
simultaneously by defining $\alpha=\alpha_\pm$. We restrict our consideration 
to values of the bulk mass parameter $a>-1$ so that $\alpha$
is real. In this case, it is clear that $\alpha_+>1$ and $\alpha_-<1$ so that
the boundary mass parameter $\alpha$ can be any real value. The case of a
massless gauge boson corresponds to $a=\alpha_-=0$. There exists a flat zero
mode, and this case has been studied extensively~\cite{gaugeb1,gaugeb2}. 
All other zero mode solutions
on the $\alpha_-$ branch are a continuous deformation of the flat mode from
$\alpha=0$ to $\infty<\alpha<1$. Notice that on this branch it is possible to
localize the massless mode on either brane. The situation on the $\alpha_+$
branch is slightly different. Boundary mass terms must be present for the zero
mode to exist on this branch, and the mode is only localized on the IR brane. 
The normalized massless mode solution for arbitrary values of $\alpha$ is
\begin{equation}
f^{0}(y)=\sqrt{\frac{2 \alpha k}{e^{2\alpha\pi k R}-1}} e^{\alpha k y}~.
\label{zeromode}
\end{equation}
The mode is localized on the UV(IR) brane for $\alpha<0$  $(\alpha>0)$. 
Interestingly, the zero mode can also be localized in the flat space limit 
$k\rightarrow 0$. The wavefunction becomes $f_0\propto e^{\pm m y}$, 
where $m$ is the bulk mass.

The massive modes are found by solving (\ref{eomy}), and are given by
\begin{equation}
f^{n}(y)=e^{ky}\Big[C_1
J_{\sqrt{1+a}}\left(\frac{m_n}{k}e^{ky}\right)+C_2
Y_{\sqrt{1+a}}\left(\frac{m_n}{k}e^{ky}\right)\Big]~,
\label{massive}
\end{equation}
and obey the following condition:
\begin{equation}
  \frac{J_{\sqrt{1+a}\pm 1}\left(\frac{m_n}{k}\right)}{Y_{\sqrt{1+a}\pm 1}
  \left(\frac{m_n}{k}\right)}=\frac{J_{\sqrt{1+a}\pm 1}\left(\frac{m_n}{k}
  e^{\pi k R}\right)}{Y_{\sqrt{1+a}\pm 1}\left(\frac{m_n}{k}e^{\pi k R}
  \right)}~.
\label{KKmasses}
\end{equation}
The masses of the Kaulza-Klein excitations are obtained from solving this 
equation. Taking the limit in the
regime $ke^{-\pi k R} \ll m_n \ll k$, we determine the mass spectrum to be
\begin{eqnarray}
       m_n\simeq\left(n\pm\frac{1}{2}\alpha_{\pm}-\frac{1}{4}\right)\,
    \pi k e^{-\pi k R}~, & n=1,2,3,\dots,
\label{KKmasses1}
\end{eqnarray}
which agrees with \cite{gbmass} for the $\alpha_-$ branch.

\subsection{Modification of the propagator}

An interesting consequence of localizing the photon is a modification 
of the propagator at high energies. It is clear from (\ref{KKmasses1}) that 
below the IR scale, only the massless photon exists and we have the 
usual massless propagator. At energies somewhat higher than this, the 
fermions will exchange massive modes and the propagator will be modified. 
The strength of the corrections depend on where the photon is localized.

To analyze these effects, we will compute the UV-UV brane Green's function. 
It is convenient to do the analysis using Poincar\'{e} coordinates, 
$z=e^{ky}/k$. The positions of the UV and IR branes in these coordinates are 
$1/k$ and $L= e^{\pi k R}/k$, respectively. Using the general
expressions in Ref.~\cite{gp2}, the Green's function in momentum space is
\begin{equation}
G(p)=- g_5^2\left( \frac{2\alpha k}{p^2}-\frac{1}{p} \frac{I_\alpha (p L)K_{\alpha+1} (p/k)+K_\alpha(p L)I_{\alpha+1} (p/k)}{I_\alpha (p L)K_{\alpha}(p/k)-K_\alpha (p L)I_\alpha(p/k)}\right)\label{green},
\end{equation}
where we have absorbed the 5D coupling $g_5$ in the propagator.
We can expand the propagator in powers of $p$ to analyze the effects of the massive modes. We will always assume that $p \ll k$. As we expect, when we expand (\ref{green}) in the regime $p \ll 1/L$ we find that the propagator is proportional to $1/p^2$ for all values of $\alpha$. The dominant exchange process comes from the massless mode and charged particles experience the ordinary $1/r$ Coulomb potential.

Now let us see what happens at high energies. The results depend on where the 
zero mode is localized in the bulk. First, consider $\alpha < 0$. Taking the 
limit $ p\gg 1/L$, we find that the propagator is given by
\begin{equation}
  G(p) \simeq -2\alpha  g_5^2 k \left( \frac{1}{p^2}+ ... -\frac{2^{2\alpha}
\Gamma(\alpha)}{\Gamma(-\alpha)}k^{2\alpha}p^{-2(\alpha+1)} \right)~.
\label{prop1}
\end{equation}
At large distance scales (compared to $1/k$) we see that the dominant 
contribution to the propagator comes from the zero mode. The corrections 
only become important at distance scales $\sim 1/k$. This is because the 
zero mode is localized on the UV brane and appears effectively pointlike 
below momentum scales of order the curvature scale $k$. However, 
for $\alpha>0$ and in the regime $p\gg 1/L$, we find
\begin{equation}
     G(p) \simeq -\frac{g_5^2 k}{2 k^2 (1-\alpha)}\left[1 
      -\frac{p^2}{4k^2(\alpha-1)(\alpha-2)}-\left(\frac{p}{2k}
    \right)^{2(\alpha-1)}\frac{\Gamma(2-\alpha)}{\Gamma(\alpha)} + ...\right]~.
\label{prop2}
\end{equation}
The dominant contribution now comes from the massive states. When the zero 
mode is localized on the IR brane, it appears to be a composite particle. 
Hence, at energies above the IR scale, the zero mode effectively disappears 
and there are only contributions from the Kaluza-Klein tower.

This correspondence will be made more explicit later where the different 
behavior exhibited by the propagator can be given a holographic 
interpretation. As we will see, the behavior of the 
propagator at high energies again depends on whether the photon is a 
composite CFT state or an elementary source field. Although the physics in 
the dual theory differs from the bulk theory, we will see that 
the result for the propagator is replicated exactly.

\section{Phenomenological implications}
We would like to examine the phenomenological implications of localizing 
Abelian gauge fields in the bulk. For the moment, we will consider a simple 
model with fermions localized on each brane and the ``photon'' residing in 
the bulk. We will describe a realistic setup within the context of the 
standard model at the end of this section in which the photon is indeed 
localized. Many of the results derived for this simple model are robust and 
will also apply in a realistic model. Moreover, as we will show later,
this simple model has a very interesting dual interpretation.

\subsection{Millicharged particles}

The ability to localize the photon allows for an interesting phenomenological
scenario in which fermions physically separated from the photon in the 
fifth dimension appear to four dimensional observers as millicharged 
particles. The existence of particles with fractional electric charge is not 
forbidden in the standard model because of the trivial commutation relations 
of the abelian group. Several theoretical models have therefore been proposed 
over the years which predict millicharged particles~\cite{ovz,holdom,flv}, 
and numerous constraints from collider experiments as well as astrophysics 
and cosmology exist for such particles~\cite{mc2,dgr}. We will see that we 
can skirt any such constraint in this model.

To begin, let us consider the 5D interaction of the U(1) field with the 
electron on the IR brane:
\begin{equation}
S  =  - \int d^5 x  \sqrt{-g}\,\Big[ g_5 \overline{\psi}(x) e^{\mu}_a \gamma^a
 A_{\mu}(x,y) \psi(x) \Big] \delta(y-\pi R)~.
\label{electron}
\end{equation}
Here, $g_5$ is the 5D coupling constant, $e^\mu_a=e^{ky}\delta^\mu_a$ is the
vielbein, and $\gamma^a$ are the ordinary flat space Dirac matrices. To examine 4D physics, we insert the
Kaluza-Klein expansion (\ref{expand}) into (\ref{electron}) and integrate over the extra
dimension. The resulting interactions between the fermion and the 
Kaluza-Klein tower are
\begin{equation}
S= - \sum_{n=0}^\infty g_5 f^{n}(\pi R) \int
d^4 x \; \overline{\psi}(x) \gamma^{\mu} A^{n}_{\mu}(x)\psi(x)~,
\label{electron2}
\end{equation}
where we have redefined the field
$\psi(x)\rightarrow e^{-\frac{3}{2}\pi kR}\psi(x)$ to canonically normalize the
fermion kinetic term. The effective 4D coupling constants can be directly 
read from (\ref{electron2}):
\begin{equation}
g_n=g_5 f^n(\pi R)~,
\end{equation}
and in particular, the electric charge is given by
\begin{equation}
e=g_5 f^{0}(\pi R)=g_5 \sqrt{\frac{2\alpha
k}{1-e^{-2\alpha \pi kR}}}~.
\label{ec}
\end{equation}
Taking the $\alpha=0$ limit correctly reproduces the result for a 5D photon 
with no bulk or boundary mass terms~\cite{gaugeb1,gaugeb2},
\begin{equation}
e = \frac{g_5}{\sqrt{\pi R}}~.
\end{equation}

Next, consider a fermion $\psi$ living on the UV brane. Following the analysis
of the electron above, we find that the coupling of $\psi$ to the photon
is now given by
\begin{equation}
    g=g_5 f^{0}(0)=g_5 \sqrt{\frac{2\alpha k}{e^{2\alpha \pi k R}-1}}~.
\label{mc}
\end{equation}
The only difference from the case of the electron is that the wavefunction is 
evaluated on the UV brane at $y=0$. Equivalently, the electric charge of 
$\psi$ can be written as $g=\epsilon e$, where $\epsilon$ is just a number. 
Then $\epsilon$ can be computed using (\ref{ec}) and (\ref{mc}):
\begin{equation}
\epsilon=\frac{f^0(0)}{f^0(\pi R)}=e^{-\alpha \pi k R}~.
\label{epsilon}
\end{equation}
The most stringent limits on $\epsilon$ are for particles with mass 
$m_\epsilon < 10^4{\rm eV}$~\cite{mc2}. For such particles, $\epsilon>10^{-14}$ has been 
ruled out. From (\ref{epsilon}), we see that this corresponds to
\begin{equation}
\alpha > \frac{14\,{\rm ln(10)}}{\pi k R}\simeq 0.9~,
\end{equation}
and since $\alpha$ can be any real value (see Eq. (\ref{alpha})) we can 
clearly produce millicharges with $\epsilon < 10^{-14}$.

It is easy to see why the fermion on the UV brane can have a much lower 
charge than the electron. The photon wavefunction is peaked on the IR brane 
and is exponentially suppressed on the UV brane for $\alpha>0$. The 
photon overlaps very weakly with the fermion on the UV brane resulting in 
the tiny coupling (\ref{epsilon}). This phenomenon is similar to what 
happens with gravity. The massless graviton also has an exponential profile 
in the bulk which explains the weakness of gravity on the IR brane.
Of course, we could have also considered the reversed situation in which the 
electron lives on the UV brane while the fermion $\psi$ lives on the 
IR brane. In this case, millicharges could be produced for $\alpha<0$.

\subsubsection{Kinetic mixing}

Usually, the fact that electric charge is quantized is thought to arise from 
grand unification. If the standard model is embedded into a larger gauge 
group, electric charge quantization is a result of the nontrival commutation 
relations of the group. However, Holdom~\cite{holdom} pointed out that the 
existence of millicharged particles is not forbidden by grand unification if 
the model contains two U(1) fields. If matter couples to both fields at 
high energies, then kinetic mixing with strength $\chi \propto e^2/(16\pi^2)$ 
will be induced by quantum corrections. Fields coupling to the second 
``shadow'' U(1) at high energies will appear as millicharges in the 
effective theory.

It is simplest to embed a massless U(1) boson into a grand unified theory
and therefore we will consider a 5D version of Holdom's model. As we will 
show in Sec. 4.3., the theory will also have a 4D dual interpretation. 
Consider two U(1) fields, $A_1^M$ and $A_2^M$ in the bulk of AdS$_5$
with $\alpha=0$. A gauge invariant operator that mixes their kinetic terms can 
be added to the action. However, to estimate the strength of this operator, 
we will assume it was generated by integrating out massive fermions 
as in \cite{holdom}.  The Lagrangian is then
\begin{equation}
{\cal L}=-\frac{1}{4 g_1^2}(F_1^{MN})^2 -\frac{1}{4 g_2^2}(F_2^{MN})^2
+ \frac{\chi}{2}\,k\,F_1^{MN}F_{2 MN}~.
\label{kinmix}
\end{equation}
where the mixing in units of $k$ is given by the dimensionless parameter 
$\chi$. Since we are assuming 
the mixing is generated perturbatively, we expect some suppression due to 
loop effects ($\chi\sim 10^{-3}$). Even in the presence of the mixing term, 
the equations of motion for $A_1^\mu$ and $A_2^\mu$ are separable. The 
solution for the zero mode is simply given by a constant. After 
compactification, the mixing of the zero modes becomes
\begin{eqnarray}
    S_{mix} & = & \chi\,k \int_0^{\pi R} dy \int d^4x~ 
    \frac{1}{2}F_1^{0\mu\nu} F^0_{2\mu\nu} + ... \nonumber \\
 & = & \chi~\pi k R \int d^4x~\frac{1}{2}F_1^{0\mu\nu}F^0_{2\mu\nu} 
+ ...~.
\end{eqnarray}
The strength of the zero mode kinetic mixing is then
\begin{equation}
\epsilon = \chi~\pi k R \sim 10^{-2}~.
\end{equation}

Following the analysis in Ref~\cite{holdom} the fields coupling to the shadow 
photon $A^{0\mu}_2$ will receive an order $\epsilon$ electric charge after 
diagonalizing the kinetic terms. A charge of $10^{-2}$ is actually quite 
constrained \cite{mc2}. Laboratory experiments rule out such a large electric 
charge for particles of mass $m_\epsilon < 100$ GeV, while astrophysical and 
cosmological considerations constrain the mass to be $m_\epsilon < 10$ TeV. 
Notice that compared to the 4D version~\cite{holdom}, 
there is an enhancement of $\pi k R \sim 36$ to the millicharge 
in this 5D model which results from integrating out the CFT.
In fact we will see that this simple 5D generalization of millicharges 
generated through kinetic mixing has a 4D dual interpretation and we will
later compare the result obtained in the bulk to the 4D theory.

\subsection{Electroweak model}
Until now we have been considering a simple model with a U(1) gauge field 
in the bulk that produces a localized zero mode. We have identified this 
mode as the photon. But the photon in the standard model is a mixture 
involving non-Abelian gauge fields. These fields would need to be similarly
localized to realize the simplest model. However, it is not clear that 
non-Abelian gauge fields can be localized in the same manner as Abelian 
gauge fields. Therefore to realize an effectively localized photon in the
standard model we will suppose that the U(1)$_Y$ gauge boson is a bulk field, 
while the SU(2)$_L$ gauge bosons and the Higgs boson are confined on the 
IR brane. To check that the proper mixing does indeed occur to produce a 
photon, $W^{\pm}$, and $Z$, consider the following 5D action
\begin{equation}
    \int d^5x \sqrt{-g} \left[ -\frac{1}{4}F_{MN}F^{MN}
    +\left(-\frac{1}{4}G^a_{\mu\nu}G^{\mu\nu a}+(D_\mu \phi)^{\dagger}
    (D^\mu \phi) -V(\phi) \right)\delta(y-\pi R)\right],
\end{equation}
where $F_{MN}$ and $G^a_{\mu\nu}$ are the field strength tensors for the 
U(1)$_Y$ and the SU(2)$_L$ gauge bosons, respectively. It is important to 
keep in mind that we are supplementing this action with the bulk and boundary 
mass terms in order to localize the U(1)$_Y$ field. The gauge covariant 
derivative is given by
\begin{eqnarray}
D_\mu & = & \partial_\mu -ig V_\mu^a(x)\frac{\sigma^a}{2} 
-i g_5 YB_\mu(x,y)~, \nonumber \\
& = &  \partial_\mu -ig V_\mu^a(x)\frac{\sigma^a}{2} 
-i g_5 Y \sum_{n=0}^\infty B^n_\mu(x)f^n(y)~,
\end{eqnarray}
where $\sigma^a$ are the Pauli matrices and $V_\mu^a, B_\mu$ are the 
SU(2)$_L$ and U(1)$_Y$ gauge bosons with $g$ and $g_5$ being their respective 
coupling to the Higgs. After decomposing the U(1)$_Y$ gauge boson and allowing the 
Higgs to acquire a vacuum expectation value:
\begin{eqnarray}
\langle \phi \rangle & = & \frac{1}{\sqrt{2}} \left( \begin{array}{c} 0 \\ v \end{array} \right),
\end{eqnarray} 
the mass Lagrangian is given by
\begin{eqnarray}
{\cal L}_m & = &
\sum_{n=1}^\infty\frac{1}{2}m_n^2
(B_\mu^{n}(x))^2+\frac{1}{2}\left(\frac{gv}{2}\right)^2\big[
(V^1_\mu(x))^2+(V^2_\mu(x))^2 \big] \nonumber \\& &
+\frac{1}{2}\left(\frac{v}{2}\right)^2\left(-g V_\mu^3(x) + g'
B^0_\mu(x)+g'\sum_{n=1}^\infty \frac{f^n(\pi R)}{f^0(\pi R)} B_\mu^n(x) \right)^2,
\end{eqnarray}
where we have used $Y=1/2$ for the hypercharge of the Higgs and 
defined the coupling of $B^0_\mu$ to be $g'= g_5 f^0(\pi R)$. 
Clearly, the $W^\pm$ bosons are defined in the standard way and the 
nontrivial mixing occurs between the $V^3$ and $B^n$ bosons. To make contact 
with the standard model, we change the basis from $(V^3, B^0)$ to $(Z, A)$ by
introducing the weak mixing angle $\theta_w$. The mass Lagrangian then
reads
\begin{equation}
{\cal L}_m = \sum_{n=1}^\infty\frac{1}{2}m_n^2
(B_\mu^{n})^2+m_W^2  W^+_\mu W^{\mu -}
+\frac{1}{2}\left(m_Z Z_\mu -
\frac{g'v}{2}\sum_{n=1}\frac{f^n(\pi R)}{f^0(\pi R)} B_\mu^n
\right)^2,
\end{equation}
where $m_W$ and $m_Z$ are defined in the usual way. We see that the photon is 
massless, but there is still mixing between the $Z$ and the Kaluza-Klein 
modes of the $B$. Since $m_n \gg m_Z$, we can diagonalize this Lagrangian
order by order. We will not do this here, but see Ref.~\cite{mp} for an 
example in flat space. The result is that the physical $Z$ contains a small 
admixture of Kaluza-Klein modes and its mass is shifted. Thus,
there is a Kaluza-Klein tower of $Z$ bosons in this setup rather than photons.
To leading order, the physical mass of the $Z$
is
\begin{equation}
  m^{(ph)2}_Z= m_Z^2\left[1-\sum_{n=1}^\infty \left( \frac{f^n(\pi R)}
  {f^0(\pi R)} \right)^2 \left(\frac{m_Z^2-m_W^2}{m_n^2}\right)\right]~,
\end{equation}
where $m_Z^2-m_W^2=(g'v/2)^2$. Only the U(1)$_Y$ gauge boson is in the bulk 
and therefore the mass corrections depend only on its coupling to the Higgs.

Although we can reproduce the standard model, we have not yet shown that any of the gauge fields, in
particular the photon, are localized in the bulk. What is actually happening is those components of
the photon and the $Z$ that come from the $B$ boson exist in the bulk and have the exponential profile given by (\ref{zeromode}).
The remaining components are confined to the IR brane. Therefore, mixing only occurs on the IR brane, and we cannot strictly define a wavefunction for the 
Standard Model gauge bosons. The most straightforward way to see that the photon and $Z$ are effectively localized is to examine their interactions with fermions. For simplicity, consider a SU(2)$_L$ singlet fermion on the IR brane. Its interaction with the $B$ boson is given by (\ref{electron}) with $A_\mu$ 
replaced with $B_\mu$. Performing a Kaluza-Klein decomposition and changing 
basis to the Standard Model gauge bosons, we find the following effective 
interaction Lagrangian:
\begin{equation}
   {\cal L}_{int} = - g' \cos \theta_w \overline{\psi}\gamma^\mu \psi A_\mu 
   + g' \sin{\theta_w}\overline{\psi}\gamma^\mu\psi Z_\mu- 
   g_5\sum_{n=1}^\infty f^n(\pi R)\overline{\psi}\gamma^\mu\psi B^n_\mu~.
\end{equation}
Note that the interaction has not yet been written in terms of the physical 
mass eigenstates of the $Z$ boson Kaluza-Klein tower. Concentrating now on the electromagnetic force, we define the electric charge of the fermion to be 
\begin{equation}
e=g' \cos \theta_w= g_5 f^0(\pi R) \cos \theta_w~.
\end{equation}
We can already guess that the electric charge of a fermion on the UV brane is 
given by $g = \epsilon \, e$, where $\epsilon$ is defined in (\ref{epsilon}), 
and our intuition is correct. Therefore, the photon and the $Z$ boson are 
effectively localized in the bulk due to the fact that their couplings to 
fermions depend on the exponential profile of $B^0_\mu$. 

We have shown that it is indeed possible to localize a U(1) gauge boson in 
a realistic context. In fact, the model we have been considering is an 
extension of the original Randall-Sundrum model (RS1)~\cite{rs}, where
we only delocalize the U(1)$_Y$ gauge boson. In the limit 
$\alpha \rightarrow \infty$, the photon and $Z$-boson are confined to the 
IR brane and we smoothly reproduce RS1. Note also that since it is not clear 
that non-Abelian fields can be localized, we have chosen to confine SU(2)$_L$
gauge bosons to the IR brane. However, it seems likely that another realistic 
model could also be constructed in which massless (flat) SU(2)$_L$ gauge 
bosons propagate in the bulk while the U(1)$_Y$ gauge boson is still localized.

\subsubsection{Electroweak constraint}

The fact that we can localize the U(1)$_Y$ gauge boson on the IR brane has 
some desirable 
phenomenological consequences. Placing massless gauge fields in the bulk 
within the context of the original RS1 model (i.e. fermions on the IR brane)
was analyzed in Ref.~\cite{gaugeb1}. One of the problems of this scenario
was that it was necessary to push the IR scale above 100 TeV in order to 
preserve the necessary condition that the bulk curvature be less than the 
5D Planck scale. Basically, this dilemma can be traced to the fact that the 
Kaluza-Klein modes couple to matter roughly 8 times stronger than does the 
zero mode. Stated another way, the problem is that the zero mode is flat in 
the bulk, while the Kaluza-Klein modes are localized near the IR brane. 
A similar problem will also occur if the U(1)$_Y$ gauge boson is in the bulk as
considered in the previous section. One way to avoid this problem is to 
delocalize the fermions as well~\cite{gp1}. However, since we can control 
the degree of localization of the U(1) bulk field with the boundary mass 
parameter, we will also be able to avoid any undesirable constraint on the 
IR scale.

At low energies, four-fermion operators will be produced by integrating out 
the Kaluza-Klein tower. The strengths of these operators will be proportional 
to a parameter $V$ defined to be
\begin{equation}
V=\sum_{n=1}^\infty \frac{g_n^2}{g_0^2} \frac{m_W^2}{m_n^2}~,
\end{equation}
where $m_n$ are the Kaluza-Klein masses.
The exchange of Kaluza-Klein modes will affect electroweak observables, and 
thus an upper limit can be placed on $V$. However, the size of $V$ depends 
on the ratio of the couplings, $g_n/g_0$. If this ratio is large, we must 
push the Kaluza-Klein mass scale and hence the IR scale to high energies to 
comply with the upper limit on $V$. 
Therefore, let us compute this ratio for arbitrary values of $\alpha$. The 
coupling of the zero mode is given by Eq. (\ref{ec}). The Kaluza-Klein tower 
couplings depend on $f^n(\pi R)$, which can be computed from (\ref{massive}) 
and are given by
\begin{equation}
f^n(\pi R)=\sqrt{2k\,\frac{Y_\alpha^2(\frac{m_n}{k})}
{Y_\alpha^2(\frac{m_n}{k})-Y_\alpha^2(\frac{m_n}{k\,e^{-\pi k R}})}}
\simeq \sqrt{2k}~.
\end{equation}
The ratio of couplings is therefore
\begin{equation}
\frac{g^n}{g^0}=\frac{f^n(\pi R)}{f^0(\pi R)}\simeq \sqrt{\frac
{1-e^{-2\alpha \pi k R}}{\alpha}}~.
\end{equation}
Taking the limit as $\alpha$ goes to zero, we find that $g^n/g^0=\sqrt{2\pi k R}\simeq 8.4$, which agrees with \cite{gaugeb1,gaugeb2}. This large coupling
forces us to push the IR scale to energies much greater than a TeV. Obviously 
if we localize the photon on the UV brane $(\alpha<0)$, the problem only 
becomes more severe. However, in the opposite limit of $\alpha >0$ when the 
photon is localized on the IR brane the ratio becomes
\begin{equation}
\frac{g^n}{g^0}\simeq\sqrt{\frac{1}{\alpha}}~.
\end{equation}
We see that for $\alpha> 1$, this ratio is actually a small number, and 
therefore the upper limit on $V$ becomes a weak constraint. This corresponds
to the gauge boson being localized near the IR brane. Clearly 
in the limit that $\alpha\rightarrow \infty$ the constraint disappears and
we recover the original RS1 model, with all gauge bosons confined on the
IR brane. 

Finally we should point out that when computing the limit on the parameter 
$V$, Ref.~\cite{gaugeb1} considered the effects of all standard model fields 
propagating in the extra dimension, whereas we have only considered a 
U(1) bulk field. Indeed, problems could arise in a model in which flat 
SU(2) gauge bosons are bulk fields even if the U(1)$_Y$ gauge boson 
is localized on the 
IR brane. Of course, in this case one would have to suppress dangerous 
four-fermion operators in other ways, such as localizing fermions in the 
bulk~\cite{gp1} and adding brane localized kinetic terms~\cite{blkt}.  

\section{Holographic interpretation}
Remarkably bulk models in a slice of AdS$_5$ can be interpreted through 
a regularized AdS/CFT correspondence as being dual to a strongly coupled 
CFT~\cite{malda,gkp,witten,arp,rz,pv}. In this modified conjecture, the fifth 
coordinate corresponds to a momentum scale in the dual 4D theory. Boundary 
values of bulk fields on the UV brane source corresponding CFT operators. 
The UV brane boundary condition implies nontrivial bulk dynamics which
in the dual theory leads to a discrete CFT spectrum, induced dynamics for 
source fields, and mixing between the source and CFT sectors. Moreover, the 
presence of the IR brane in the bulk theory is holographically interpreted 
as a deformation of the CFT, with conformal invariance spontaneously broken 
in the IR. 

We will therefore be interested in giving the holographic interpretation of 
the localized U(1) gauge field in a slice of AdS$_5$. First, however, it 
will be useful to review several aspects regarding the duality of massless 
gauge fields (flat-profile zero mode) that have been discussed in the 
literature~\cite{arp,ad}. The dual theory is a strongly coupled CFT gauged by 
an external source field. The theory contains a massless spin one field 
(``photon'') that is a mixture of the source and CFT fields. The situation 
is somewhat analogous to $\gamma -\rho$ mixing in QCD; however, in this case 
there is strong mixing between the source and CFT because the zero mode is 
flat in the bulk. The 4D coupling constant resulting from the overlap 
integral of the zero mode wavefunctions is interpreted as a logarithmically 
running coupling constant evaluated at the IR scale. Also, corrections to 
source propagator induced by CFT loops are seen in the bulk theory as 
contributions of the Kaluza-Klein tower to the UV-UV brane propagator. 

The fact that the massless mode can be localized anywhere in the bulk has 
interesting consequences for the dual picture. As we will see, when the mode 
is localized on the UV brane, the photon eigenstate in the dual theory is 
primarily composed of the source field. To continue the analogy with QCD, 
in this case the photon eigenstate is mostly the elementary QED photon, with 
a tiny admixture of the QCD composite state (``$\rho$'').  In fact this case
mimics quite well the situation in QCD. However, if the massless mode is 
localized on the IR brane, the situation is reversed. The photon eigenstate
is then mostly a massless composite state while the source field (``QED 
photon'') becomes massive and contributes very little to the mass eigenstate.
The dual picture is in fact qualitatively similar to that derived for 
localized fermions~\cite{cp} and localized gravitons~\cite{gpp}.

In addition to seeing the different aspects of elementarity and 
compositeness, it will also be interesting to give the dual interpretation of 
the bulk couplings and interactions with fermions living on 
the UV brane. To this end, we will examine the two-point function of the 
dual CFT operator $J$ and describe its behavior in different energy
regimes. The analysis will be done using Poincar{\'e} coordinates, and 
the position of the UV and IR branes will now be at $z=L_0$ and $z=L_1$, 
respectively.

Our starting point is the 5D homogeneous equation of motion (written in 
momentum space):
\begin{equation}
\left[ \partial_5^2-\frac{1}{z}\partial_5 -\left(p^2+\frac{a}{z^2}
\right)\right] A_\mu(p,z)=0~.
\end{equation}
The solution is given by
\begin{equation}
A_{\mu}(p,z)=C_{\mu}(p) pz \left[K_{
\alpha -1} (pz)+b \, I_{\alpha-1}(pz)\right]\; \equiv \;C_{\mu}(p) A(z)~,
\label{bulkeom}
\end{equation}
where $\alpha=1\pm \sqrt{1+a}$ is the boundary mass parameter and the 
coefficient $b$ is determined from the IR boundary condition,
\begin{equation}
   b = \frac{K_{\alpha}(p L_1)}{I_{\alpha}(p L_1)}~.
\end{equation}
Next, we evaluate the bulk action for an arbitrary UV boundary condition for 
the gauge field, $A_{\mu}(x,L_0) = {\widehat A}_{\mu}(x)$. The regularized 
correspondence states that in the dual theory, the boundary value acts as a 
source for a CFT current $J$:
\begin{equation}
\Big\langle {\rm exp}\left(-\int d^4x \widehat A_{\mu}J^{\mu}\right)
\Big\rangle_{\rm{CFT}}={\rm exp}[-\Gamma({\widehat {A}})]~.
\label{adscfteq}
\end{equation}
Here the LHS is the generating functional of CFT correlation functions. The 
effective action is
\begin{eqnarray}
\Gamma({\widehat A}) & = &-\frac{1}{2 g_5^2}\int d^4x \; \eta^{\mu\nu}
 \; \left( \frac{1}{k z} A_{\mu}(x,z)\partial_5 A_{\nu}(x,z)\right.
~\nonumber\\
&&\qquad\qquad\qquad\qquad\qquad\qquad-\left.\frac{1}{(k z)^3} 
\alpha \,k (k L_0) A_{\mu}(x,z) 
A_{\nu}(x,z) \right) \bigg\vert_{z = L_0}~, \nonumber \\  
& = & -\frac{1}{2 g_5^2k}\frac{1}{L_0} \int \frac{d^4p}{(2\pi)^4}\; \eta^{\mu\nu}\;{\widehat
A}_{\mu}(p){\widehat A}_{\nu}(-p) \left( \frac{\partial_5 A}{A}-\frac{\alpha}{L_0} \,
\right) \bigg\vert_{z = L_0}~, \nonumber \\ 
&=& -\frac{1}{2}\int \frac{d^4p}{(2\pi)^4}\; \eta^{\mu\nu}\;
{\widehat A}_{\mu}(p) \Sigma(p) {\widehat A}_{\nu}(-p)~,
\label{effaction}
\end{eqnarray}
where
\begin{equation}
\Sigma(p)=-\frac{1}{g_5^2 k}\frac{p}{L_0}
\frac{K_{\alpha}(p L_0)I_{\alpha}(p L_1)-I_{\alpha}(p L_0)K_{\alpha}(p L_1)}{K_{\alpha -1}(p L_0)I_{\alpha}(p L_1)+I_{\alpha-1}(p L_0)K_{\alpha}(p L_1)}~.
\label{sigma} 
\end{equation}
Differentiating twice with respect to the source $\widehat{A}_\mu$ yields the 
$\langle JJ \rangle$ correlator which is contained (up to the overall tensor 
structure) in $\Sigma(p)$.

First note that the expression for $\Sigma(p)$ is only valid for momentum 
scales below the UV cutoff, $p < 1/L_0$. However, we can examine the behavior 
of the correlator at energies above and below the IR scale by expanding
in the momentum $p$. At high energy $ p L_1 \gg 1$ we can essentially neglect 
the presence of the IR brane;  the dual theory is conformal in this regime and 
the leading nonanalytic piece results from CFT dynamics. In the 5D picture, 
this behavior results from the exchange of Kaluza-Klein modes at high energies. Also present are terms analytic in $p$, which we interpret as source dynamics. At low energies, $p L_1 \ll 1$, we no longer have the Kaluza-Klein tower in the bulk and the conformal piece vanishes. As we will see, the interesting features of this behavior in these two distinct regimes depends heavily on where the gauge field is localized in the extra dimension.

\subsection{$\alpha_-$ branch holography}
Consider first the correlator $\Sigma(p)$ on the $\alpha_-$ branch, where $-\infty< \alpha < 1$. At momentum scales far above the IR, $p L_1 \gg 1$, but below the UV, $p L_0 \ll 1$, 
we find for noninteger $\alpha$
\begin{equation}
\Sigma(p) \simeq  \frac{1}{g_5^2 k}\frac{1}{2\alpha}\left[p^2 \, 
+\, \cdots\,  + \frac{2^{2\alpha} \;\Gamma(\alpha)}{\Gamma(-\alpha)}
L_0^{-2\alpha} p^{2-2\alpha}\, +\,  \cdots \right],
\label{sigma1}
\end{equation}
where we have included only the leading terms in the expansion. Note that 
for integer $\alpha$ the nonanalytic terms will instead be logarithmic. 
We will not treat integer $\alpha$ here, since the analysis is qualitatively 
similar to the noninteger case.
As we expect, the conformal (nonanalytic) term is present and the effects 
of the IR brane are irrelevant. We can isolate the two-point function 
$\langle J J \rangle$ by defining a rescaled source field 
$\widehat{A}_\mu \rightarrow L_0^{\alpha} \widehat{A}_\mu$ in the effective 
action (\ref{effaction}) and taking the limit $L_0 \rightarrow 0$. In this 
limit, terms analytic in $p$ that are divergent can be cancelled by adding 
appropriate counterterms in the boundary action. This is the customary 
prescription used in the stringy correspondence to make contact with the 
CFT on the AdS boundary. The correlator is then given by
\begin{equation}
  \langle JJ \rangle (p) = \frac{1}{g_5^2 k}
   \frac{2^{2\alpha} \;\Gamma(\alpha)}{2 \alpha\, \Gamma(-\alpha)}
   p^{2(1-\alpha)}~.
   \label{jj1}
\end{equation}
The scaling dimension of $J$ can be found by Fourier 
transforming this term, and is given by
\begin{equation}
\Delta_J=3-\alpha,
\label{dimJ-}
\end{equation}
which can be as low as 2 when $\alpha=1$. This deviates from the canonical
dimension of $J$, namely $[J]=3$, as occurs for the case of a flat zero 
mode ($\alpha=0$) and leads to an anomalous dimension $-\alpha$. The 
leading analytic piece is interpreted as a kinetic term for the source field 
that is induced via interactions with the CFT. The absence of a constant term 
in $\Sigma(p)$ tells us that the source field is massless in the 4D theory. 

Previously we considered the interaction of the bulk gauge field with 
fermions on the UV brane, and we can directly include this interaction in 
the dual theory. Hence the Lagrangian of our dual theory below the cutoff scale
$\Lambda=1/L_0 \sim k$ is given by
\begin{equation}
  {\cal L}_{4D} = -\frac{1}{4}Z_0 F_{\mu\nu}^2 + 
   \Lambda^{\alpha} A_\mu J^\mu + \overline{\psi}\gamma^\mu 
   A_\mu \psi + {\cal L}_{CFT}~,
\label{4dlag}
\end{equation}
where $Z_0$ is a dimensionless coupling. In fact from 
(\ref{sigma1}) one can read off $Z_0= -1/(2g_5^2 k\alpha)$. Because of the 
anomalous dimension of $J$ (\ref{dimJ-}), it is clear that the coupling of 
the source to the CFT 
current is relevant for positive $\alpha$, marginal for $\alpha=0$, and 
irrelevant for negative $\alpha$. Thus for negative $\alpha$ we can neglect
the source coupling to the CFT and the mass eigenstate of the photon is 
primarily composed of the source field. Instead for the marginal or relevant
couplings the mixing between the source and CFT sector will result in a part
elementary and part composite photon eigenstate.

Below the IR scale, we expect that conformal invariance will be broken. 
Physically, we have integrated out the massive CFT degrees of freedom at the 
IR scale. This will induce an extra contribution to the kinetic term of the 
photon. We can see this effect exactly by calculating $\Sigma(p)$ for 
energies $p\ll 1/L_1$:
\begin{equation}
\Sigma(p)_{IR} \simeq \frac{1}{g_5^2 k}
\frac{1}{2\alpha}\big[ 1- (L_1/L_0)^{2\alpha}\big]p^2 + \dots~.
\label{sigIR}
\end{equation}
The disappearance of the nonanalytic piece signals the breaking of conformal 
invariance in the IR. Moreover, we now see a contribution to the kinetic term 
arising from integrating out CFT dynamics. This suggests that one can define 
a running wavefunction $Z(\mu)$ where 
\begin{equation}
    Z(1/L_1) = \frac{1}{g_5^2 k}\frac{1}{2\alpha}((L_1 \Lambda)^{2\alpha}-1)~.
    \label{Z}
\end{equation}
Canonically normalizing the 
Lagrangian (\ref{4dlag}), we find the low energy effective coupling of the 
source field to matter is given by
\begin{equation}
   g=\frac{1}{\sqrt{Z(1/L_1)}}=g_5\sqrt{\frac{2\alpha k}
   {e^{2\alpha \pi k R}-1}}~.
\end{equation}
This precisely matches the bulk calculation for the effective 4D charge 
(\ref{mc}) for fermions on the UV brane when $L_0=1/k$. 
The strength of the coupling depends on 
whether or not the photon is mostly elementary ($\alpha<0$) or composite 
$(0<\alpha < 1)$. 

Using the running wavefunction $Z(\mu)$ (obtained from (\ref{sigma1}))
we can write down a renormalization group equation which encodes the mixing 
behavior as was done for fermions~\cite{cp}~\footnote{We 
thank R. Contino for helpful discussions on the fermion case.}. Define the 
dimensionless coupling 
$\omega(\mu)= 1/\sqrt{Z(\mu)} (\mu/\Lambda)^{-\alpha}$, then we obtain
\begin{equation}
\mu\frac{d\omega}{d\mu}=-\alpha \omega+ c \frac{N}{16\pi^2} \omega^3~,
\label{rge}
\end{equation}
where $1/(g_5^2 k) = N/(16\pi^2)$ and $c$ is a constant. The second term 
in (\ref{rge}) arises from the
CFT contribution to $Z_0$. When $0<\alpha <1$, the constant $c>0$ and there
is a fixed point $\omega_\ast\sim 4\pi \sqrt{\alpha/(cN)}$, corresponding to 
the fact that the $L_1/L_0$ term in 
(\ref{sigIR}) dominates the kinetic term at low energies. This corresponds to
nonnegligible mixing between the source and CFT sector. On the other hand
when $\alpha<0$ we can neglect the second term in (\ref{rge}) and the solution
corresponds to the simple scaling behavior $\omega \sim 4\pi\sqrt{-\alpha/N}
(\mu/\Lambda)^{-\alpha}$, where we have matched to the low energy value 
(\ref{Z}). Clearly at low energies $(\mu\ll \Lambda\sim k)$ the mixing will 
diminish and the contribution from the CFT sector is not important.

\subsubsection{Interactions with external fermions}

\begin{figure}
\centerline{\includegraphics[width=1\textwidth]{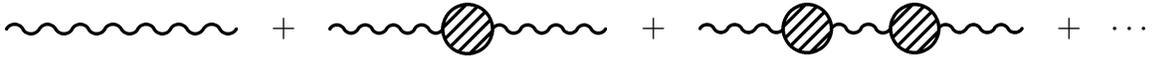}}
\caption{The Feynman diagrams in the 4D dual theory responsible for the
corrections to the propagator. The source field $A_\mu \,$, interacts with 
the CFT contribution, indicated by the blob.}
\label{cftfig}
\end{figure}

The mixing between the source and CFT sector has important effects on 
fermionic interactions. Because the source couples directly to external 
fermions, interactions are mediated through the source propagator. It is 
important to realize that in the dual 4D theory the physical photon is a 
combination of source and 
CFT states. We will therefore examine corrections to the propagator which 
arise from insertions of CFT correlators as shown in Fig.~\ref{cftfig}. 
The infinite series of Feynman diagrams can easily be summed in the 
following way:
\begin{eqnarray}
G(p) & = & \frac{1}{Z_0 \, p^2}\left[ 1- \Lambda^{2\alpha} 
\frac{\langle JJ \rangle (p)}
{Z_0 \, p^2} +  \left(\Lambda^{2\alpha}\frac{\langle JJ \rangle (p)}{Z_0 \, p^2}\right)^2 - 
\dots\right]~,\nonumber \\
 & = &   \frac{1}{Z_0\,p^2 + \Lambda^{2\alpha} \langle JJ \rangle (p)}\nonumber~, \\
& = & \frac{1}{\Sigma(p)}~.
\label{dualprop}
\end{eqnarray}
This is what we would expect from examining the bulk effective action 
(\ref{effaction}), and provides a nontrival check of the holographic 
correspondence.

How can we physically see what is happening in the 4D theory? Let us first consider $\alpha< 0$, in which case the coupling of the source field to the CFT is irrelevant. In this case, the photon is mostly comprised of the source field. Thus, the correction arising from the CFT is small, and we can expand the denominator in (\ref{dualprop}) to find 
\begin{equation}
    G(p) \simeq 
  \frac{1}{Z_0\, p^2}-\Lambda^{2\alpha}\frac{\langle JJ \rangle (p)}{Z_0^2 \, p^4}~.
\label{dualprop1}
\end{equation}
By inserting $\langle J J \rangle$ (\ref{jj1}) into (\ref{dualprop1}) and 
using (\ref{Z}), we recover the result previously obtained in the bulk 
gravity calculation (\ref{prop1}). Contributions from the CFT are only 
important in the UV. As we flow to the IR, the coupling between the source 
and the CFT becomes negligible and the interaction is mediated solely by the 
source field. The marginal case ($\alpha=0$) is special and can be treated 
in a similar fashion as was done for gravity~\cite{gpp}.

For relevant couplings ($\alpha>0$), the CFT contribution dominates, and 
we can neglect the source contribution:
\begin{equation}
   G(p)  \simeq \frac{1}{\Lambda^{2\alpha}\langle JJ \rangle (p)}~.
\label{dualprop2}
\end{equation} 
Strong mixing between the source and CFT fields combine to produce the 
massless photon for $0<\alpha <1$. It therefore makes sense that the CFT 
contribution is indeed important to the interaction between external fermions. 
Again rewriting the couplings in terms of the 4D charge, it is easy to verify 
that (\ref{dualprop2}) matches the bulk gravity calculation (\ref{prop2}) 
exactly.

It is remarkable that both the couplings and the propagators of the 4D and 
5D theories match precisely, and complement those found for the Newtonian 
potential~\cite{gpp}. The tree level effects in the classical gravity 
theory are realized as first order corrections in the dual CFT. Although the 
dynamics of the dual theory are a mystery, we can compute quantum effects 
directly using holography. 

\subsection{$\alpha_+$ branch holography}
We now consider the $\alpha_+$ branch ($\alpha>1$), in which the photon is 
always localized on the IR brane. Expanding the two point function in the 
regime $1/L_1\rightarrow 0$ and $1/L_0\rightarrow \infty$, we find
\begin{equation}
   \Sigma(p)  \simeq  -\frac{1}{g_5^2 k}  \Big[ 2(\alpha-1)L_0^{-2}  +
  \frac{1}{2(\alpha-2)}p^2+\cdots 
      +\frac{\Gamma(2-\alpha)}{2^{2\alpha-3}
   \Gamma(\alpha-1)}L_0^{2\alpha-4} p^{2(\alpha-1)}+ \cdots \Big]~.
\label{sigma2}
\end{equation}
Following the same renormalization procedure as in the previous section, 
we can extract the $\langle J J \rangle$ correlator from (\ref{sigma2}):
\begin{equation}
 \langle J J \rangle = -\frac{1}{g_5^2 k}\frac{\Gamma(2-\alpha)}{2^{2\alpha-3}
   \Gamma(\alpha-1)}p^{2(\alpha-1)}~.
\end{equation}
The scaling dimension of $J$ is therefore
\begin{equation}
\Delta_J=\alpha +1~.
\end{equation}
In this case the anomalous dimension is $\alpha-2$.
It also appears that the source field has become massive, as indicated by 
the leading constant analytic piece. 

To ascertain what happened to the massless particle, let us expand 
$\Sigma(p)$ in the low energy regime, $p L_1 \ll 1$:
\begin{equation}
\Sigma(p)_{IR}=-\frac{1}{g_5^2 k}  \left[2  (\alpha-1) L_0^{-2}  +
\frac{1}{2 (\alpha-2)} p^2 - 8\alpha(\alpha-1)^2 \frac{L_0^{2\alpha-4}}
{ L_1^{2\alpha}}\frac{1}{p^2} +\dots\right]~.
\label{sigma3}
\end{equation}
We notice the appearance of a pole at $p^2=0$. This implies that the photon is 
primarily a CFT bound state. Similar massless bound states have also been 
found for bulk scalars~\cite{rz,gp3}, fermions~\cite{cp}, and 
gravitons~\cite{gpp}. Notice what happens 
as we transition from the $\alpha_-$ to the $\alpha_+$ branch. The source, 
which was massless on the $\alpha_-$ branch, obtains a mass at the same 
point that the CFT produces a composite massless vector field.

We can now write the Lagrangian of the dual theory as
\begin{equation}
{\cal L}_{4D}=-\frac{1}{4} {\widetilde Z}_0 F^2_{\mu\nu}
-\frac{1}{2}m_0^2A_\mu A^\mu
 +\frac{1}{\Lambda^{\alpha-2}} A_\mu J^\mu+ 
\overline{\psi}\gamma^\mu A_\mu \psi~,
\end{equation}
where ${\widetilde Z}_0$ is a dimensionless parameter and $m_0$ is a mass
parameter of order the curvature scale. In fact from (\ref{sigma2}) we can 
read off that ${\widetilde Z}_0=1/(2g_5^2 k(\alpha-2))$ and 
$m_0^2=2\Lambda^2(\alpha-1)/g_5^2k$. 
In contrast to the $\alpha_-$ branch, the source-CFT interaction remains in the IR due to the fact that CFT contains a massless composite particle. Because the source has become heavy, it effectively decouples at low energy from matter. The photon propagates through its interactions with the source field. Therefore, the dominant contribution to the propagator is given by a single insertion of the $\langle JJ \rangle$ correlator. Noting that at large distances, the correlator is given by
\begin{equation}
\langle JJ \rangle (p) \simeq \frac{1}{g_5^2 k} 8 \alpha (\alpha-1)^2 L_1^{-2\alpha}\frac{1}{p^2}~,
\end{equation}
we can calculate the propagator in a straightforward manner:
\begin{eqnarray}
G(p) & \simeq &  \frac{1}{{\widetilde Z}_0 p^2+m_0^2}
    \left(\frac{ \langle JJ \rangle (p)}{\Lambda^{2(\alpha-2)}} \right) \frac{1}{{\widetilde Z}_0 p^2+m_0^2}~, 
    \nonumber \\ & \simeq & \frac{\langle JJ \rangle (p)}{m_0^4\Lambda^{2(\alpha-2)}}~, \nonumber \\
     & = & 2 \alpha g_5^2 k (\Lambda L_1)^{-2\alpha} \frac{1}{p^2}~.
\end{eqnarray}
In the second line we have neglected the $p^2$ part in the source propagator 
which is valid for the momentum scales we are considering. Taking the 
nonrelativistic limit, we see that the Coulomb potential emerges at low energy:
\begin{eqnarray}
V(r) & = & \int \frac{d^3 p}{(2\pi)^3} e^{i {\bf p} \cdot {\bf x}} 
  G({\bf p})~,\nonumber \\
     & = & 2 \alpha g_5^2 k (\Lambda L_1)^{-2\alpha}\int \frac{d^3p}{(2\pi)^3} 
     e^{i{\mathbf p} \cdot {\mathbf x}}\frac{1}{{\mathbf p}^2}~,\nonumber\\
     &=&\frac{g^2}{4\pi r}~,
\end{eqnarray}
where we have written the electric charge as defined in (\ref{mc}) when the 
UV brane is at $L_0=1/k$. Thus, we see that in the dual 4D theory the 
millicharge arises because the UV fermion must now couple to a composite 
photon. This coupling to the CFT vector current (with large anomalous 
dimension) can only occur via the massive source field.

Above the IR scale, it is clear from (\ref{sigma2}) that although the source 
remains massive, the pole disappears, indicating the absence of a massless 
particle. This is what we would expect from the bulk calculation of the 
propagator, in which the $1/p^2$ term vanishes at high energies. Again, the 
$\langle JJ \rangle$ correlator yields the dominant contribution to the 
propagator:
\begin{eqnarray}
   G(p) & \simeq & \frac{ \langle JJ \rangle (p)}{m_0^4 
 \Lambda^{2(\alpha-2)}}~, \nonumber \\
  & =& -g_5^2 k\frac{\Gamma(2-\alpha)}{2^{2\alpha-1}(\alpha-1)
\Gamma(\alpha)}\Lambda^{-2\alpha} p^{2(\alpha-1)}~.
\end{eqnarray}
which agrees with the result obtained in the gravity dual (\ref{prop2}). 

Therefore, on the $\alpha_+$ branch, CFT dynamics produces a nontrivial
effective interaction at high energies. However, large distance interactions 
are mediated by a massless vector particle which is primarily a CFT bound 
state. As we transition to the IR, the photon emerges from the CFT and the 
standard low energy theory is reproduced. This emergent photon behavior is 
similar to the emergent gravity behavior obtained in Ref.~\cite{gpp}.

\subsection{Kinetic mixing dual interpretation}

Let us consider the dual interpretation of the bulk kinetic mixing discussed 
in Sec. 3.1.1. If we consider the fields $A_1^M$ and $A_2^M$ in our bulk 
theory, there will exist corresponding operators $J_1$ and $J_2$ in the dual 
CFT. As we will show, if there is kinetic mixing in the bulk, a corresponding 
kinetic mixing will be induced in the dual theory. Hence this theory can be 
considered analogous to the mechanism in Ref~\cite{holdom}, but coupled to 
a strongly interacting sector.

The bulk theory is governed by the 5D Lagrangian (\ref{kinmix}). Again 
for simplicity we assume the bulk fields are massless to ensure that both 
U(1)'s can be embedded in a grand unified theory. Using the bulk solution to 
the homogeneous equations of motion for $A_1^\mu$ and $A_2^\mu$, which is 
given (up to an overall constant) in (\ref{bulkeom}) with $\alpha=0$, we can 
calculate the effective gravity action. Here we only consider the portion 
contributing to the $\langle J_1 J_2 \rangle$ correlator, which is 
given by
\begin{eqnarray}
     \Gamma(\widehat{A}_1,\widehat{A}_2) & = & \frac{\chi k}{2}\int 
     \frac{d^4 p}{(2\pi)^4}~
     \eta_{\mu\nu}\widehat{A}_1^\mu \left( \frac{\partial_5 A_1}{A_1} 
    + \frac{\partial_5 A_2}{A_2}\right) \widehat{A}_2^\nu~, \nonumber \\
& = & \frac{1}{2}\int\frac{d^4 p}{(2\pi)^4}~\eta_{\mu\nu} \widehat{A}_1^\mu 
\Sigma_{12}(p) \widehat{A}_2^\nu~, 
\end{eqnarray}
with $\Sigma_{12}(p)$ defined by
\begin{equation}
\Sigma_{12}(p)=-\chi  \frac{p}{L_0}
\frac{K_{0}(p L_0)I_{0}(p L_1)-I_{0}(p L_0)K_{0}(p L_1)}{K_{1}(p L_0)
I_{0}(p L_1)+I_{1}(p L_0)K_{0}(p L_1)}~.
\label{sigma12}
\end{equation}
Note that the calculation is similar to that performed in Ref.~\cite{arp} for 
a single massless gauge field. The correlator is found by differentiating 
with respect to $\widehat{A}_1^\mu$ and $\widehat{A}_2^\mu$

The existence of a nonvanishing  $\langle J_1 J_2 \rangle$ correlator implies 
that a kinetic mixing for the source fields will receive corrections from
CFT loops. Expanding $\Sigma_{12}(p)$ for low momentum scales, $pL_1 \ll 1$:
\begin{equation}
 \Sigma_{12}(p)\simeq \chi\log(L_1/L_0)~p^2+\dots~,
\end{equation}
we can read off the strength of the mixing as 
\begin{equation}
\epsilon = \chi \log(L_1/L_0) = \chi\,\pi k R \sim 10^{-2}~,
\end{equation}
which agrees identically with the bulk calculation. Thus, fermions coupling 
to the shadow U(1) will acquire an electric charge of order $10^{-2}e$ from
the CFT sector.

\section{Conclusion}

The zero modes of 5D U(1) gauge fields in a slice of AdS can
be localized anywhere in the bulk. We employed a 5D St\"uckelberg mechanism 
in order to maintain gauge invariance even though bulk and boundary masses
are added to the 5D action. A simple model of millicharged particles
can then be constructed, which allows fermions to have arbitrarily small 
electric charge. In the electroweak sector only the U(1)$_Y$ gauge boson can 
be localized, but leads to the effective localization of electric charge. We 
have also showed that stringent electroweak constraints on the IR scale 
from bulk Abelian gauge fields can be avoided by localizing the U(1)$_Y$ 
gauge boson close to the IR brane.

We have also presented the detailed holographic interpretation of the
localized U(1) gauge field in the warped 5D bulk. When the zero mode
is localized near the UV (IR) brane the photon eigenstate in the
4D dual theory is predominantly an elementary (composite) state. The 
composite photon is an example of emergent behavior because above the
compositeness scale (at short distances) the photon disappears.
We also verified that when the CFT has a massless pole (corresponding to 
the composite photon) the source field receives a mass of order the 
curvature scale. In this way the dual theory is consistent and there 
is always only one massless state. Furthermore in the dual theory, 
millicharged particles are understood as arising from fermions which
couple to vector currents with large anomalous dimensions. The electric 
charge is then proportional to this coupling and can be arbitrarily small.
Thus, all the physics of localized Abelian gauge fields in the warped bulk
can be given a purely 4D holographic description.

\section*{Acknowledgements}
This work was supported in part by a Department of Energy grant 
DE-FG02-94ER40823 at the University of Minnesota, a grant from the Office 
of the Dean of the Graduate School of the University of Minnesota, and an 
award from Research Corporation.

\end{document}